\documentclass[epj]{webofc}
\usepackage[varg]{txfonts}   
\usepackage{amssymb,amsmath,epsfig,bm,pifont}
\wocname{EPJ Web of Conferences}
\woctitle{CONF12}
\begin{document}
\selectlanguage{english}
\title{Improved estimates of the pion-photon transition form factor
       in the $(\mathbf{1\leq Q^2\leq5})$~GeV$^\mathbf{2}$ range and
       their theoretical uncertainties}
%
%

\author{S.~V.~Mikhailov\inst{1}\fnsep\thanks{\email{mikhs@theor.jinr.ru}} \and
        A.~V.~Pimikov\inst{1,2}\fnsep\thanks{\email{pimikov@mail.ru}} \and
        N.~G.~Stefanis\inst{3}\fnsep\thanks{\email{stefanis@tp2.ruhr-uni-bochum.de}}
}

\institute{Bogoliubov Laboratory of Theoretical Physics, JINR,
           141980 Dubna, Russia
\and
           Institute of Modern Physics, Chinese Academy of Sciences,
           Lanzhou, 730000, P. R. China
\and
           Institut f\"{u}r Theoretische Physik II,
           Ruhr-Universit\"{a}t Bochum,
           D-44780 Bochum, Germany
}

\abstract{%
We consider the pion-photon transition form factor at low to
intermediate spacelike momenta within the theoretical framework
of light-cone sum rules.
We derive predictions which take into account all currently
known contributions stemming from QCD perturbation theory up to
the next-to-next-to-leading order (NNLO) and by including all twist
terms up to order six.
In order to enable a more detailed comparison with forthcoming
high-precision data, we also estimate the main systematic theoretical
uncertainties, stemming from various sources, and discuss their
influence on the calculations --- in particular the dominant one
related to the still uncalculated part of the NNLO contribution.
The analysis addresses, in broad terms, also the role of the
twist-two pion distribution amplitude derived with different
approaches.
}
\maketitle
\section{Introduction}
\label{sec:intro}
Collinear factorization
\cite{Efremov:1979qk, Efremov:1978rn, Lepage:1979zb, Lepage:1980fj}
provides a firm foundation for the application of Quantum
Chromodynamics (QCD) to the hard exclusive reactions of hadrons.
This makes it possible to calculate partonic subprocesses in
a controllable way within QCD perturbation theory.
In addition, one needs a reliable framework to include higher twist
contributions which represent nonperturbative power corrections.
The binding effects ensuing from the confining quark-gluon dynamics
are factorized out and have to be determined by other means, e.g.,
from experiment, lattice simulations, or nonperturbative models.

Consider, for example, the process
$\gamma^*(q_{1}^{2})\gamma^*(q_{2}^{2})\to \pi^0$,
with $q_{1}^{2}=-Q^2$ and
$q_{2}^{2}=-q^2$.
Here the leading-twist two transition form factor (TFF) for two
highly off-shell photons can be written in factorized form at the
momentum scale $\mu_{\rm F}^2$ in the form \cite{Brodsky:1981rp}
\begin{eqnarray}
  F^{\gamma^{*}\gamma^{*}\pi^0}\left(Q^2,q^2;\mu_{\rm F}^2\right)
=
  T\left(Q^2,q^2;\mu_{\rm F}^2;x\right)
\otimes
  \varphi_{\pi}^{(2)}\left(x,\mu_{\rm F}^2\right) \, ,
\label{eq:fact-TFF}
\end{eqnarray}
where
$\varphi_{\pi}^{(2)}\left(x,\mu_{\rm F}^2\right)$
is the pion distribution amplitude (DA) of leading-twist two and
$\otimes \equiv \int_{0}^{1} dx$.
The pion DA has a nonperturbative origin and interpolates between the
partonic degrees of freedom of QCD (quarks and gluons) and the
pion bound state
\begin{eqnarray}
  \langle 0| \bar{q}(z) \gamma_\mu\gamma_5 [z,0] q(0)
           | \pi(P)
  \rangle|_{z^{2}=0}
=
  if_\pi P_\mu \int_{0}^{1} dx e^{i x (z\cdot P)}
  \varphi_{\pi}^{(2)} \left(x,\mu_{\rm F}^{2}\right) \, ,
\label{eq:pion-DA}
\end{eqnarray}
where we have employed the lightcone gauge $A^+=0$ in order to
reduce the gauge link
$
 [z,0]
=
 \mathcal{P}\text{exp}
 \left[-ig\int_{0}^{z}dy_{\mu}A_{a}^{\mu}(y)t_{a}\right]
$
to unity.
The DA
$\varphi_{\pi}^{(2)} \left(x,\mu^{2}\right)$
describes the partition of longitudinal-momentum fractions
between the valence quark
($x_q=x=(k^0+k^3)/(P^0+P^3)=k^+/P^+$)
and the valence antiquark
($x_{\bar{q}}=1-x\equiv \bar{x}$)
at the reference scale $\mu$
subject to the normalization condition
$
 \int_{0}^{1}dx\varphi_{\pi}^{(2)}(x, \mu^{2})
=
 1
$.

The virtue of the collinear factorization is that the hard-scattering
amplitude $T\left(Q^2,q^2;\mu_{\rm F}^2;x\right)$ can be cast in the
form of a power-series expansion in terms of the running strong
coupling
$a_s\equiv \alpha_{s}(\mu_{\rm R}^2)/4\pi$:
\begin{equation}
  T\left(Q^2,q^2;\mu_{\rm F}^2;x\right)
=
  T_\text{LO} + a_s~T_\text{ NLO} + a_{s}^2~T_\text{NNLO} + \ldots
\label{eq:hard-scat-ampl}
\end{equation}
becoming calculable within perturbative QCD.
Note that the so-called default scale setting has been adopted, i.e.,
the renormalization scale $\mu_{\rm R}$ and the
factorization scale $\mu_{\rm F}$ have been identified, so that
$a_s\left(\mu_{\rm R}^{2}\right)=a_s\left(\mu_{\rm F}^{2}\right)$.
Moreover, we are using here and below the following convenient
abbreviations:
LO (leading order),
NLO (next-to-leading order),
NNLO (next-to-next-to-leading order).
These contributions will be denoted in terms of the labels
(0), (1), and (2), respectively.
The NLO term is completely known
\cite{Melic:2002ij, Bakulev:2002uc},
whereas the calculation of the NNLO contribution has not been
completed yet.
As we will discuss later in more detail, the uncalculated part
entails the strongest theoretical uncertainty
(see \cite{Mikhailov:2016klg} for a deeper exploration of this issue).

Adopting particular models for the pion DA, one can safely compute
at the leading-twist two level $F^{\gamma^{*}\gamma^{*}\pi^0}$
using the above expansion and taking into account the
Efremov-Radyushkin-Brodsky-Lepage (ERBL) evolution equation
\cite{Efremov:1978rn, Lepage:1979zb, Lepage:1980fj}
in order to connect the result obtained at the initial scale
$\mu_{0}$ to any higher momentum value.
However, in real experiments one of the photons is almost real,
meaning that its virtuality $q_{2}^{2}\gtrsim 0$ is so small that
one needs to include its hadronic large-distance content.
As a result, perturbative QCD cannot be reliably applied to the
calculation of the
$F^{\gamma^{*}\gamma\pi^0}\left(Q^2,q^2\gtrsim 0\right)$
TFF.

A convenient framework to calculate the TFF for $q^2\to 0$, is
provided by the method of light-cone sum rules (LCSR), developed in
\cite{Balitsky:1989ry, Khodjamirian:1997tk},
which relies upon local duality in the vector channel in terms of a
dispersion relation.
The key ingredient of this framework is the spectral density
\begin{equation}
  \bar{\rho}(Q^2,x)
=
  (Q^2+s) \rho^{\text{pert}}(Q^2,s) \, ,
  \label{eq:rho-bar}
\end{equation}
where
\begin{eqnarray}
 \label{eq:rho}
  \rho^{\text{pert}}(Q^{2},s)
=
  \frac{1}{\pi} {\rm Im}\left[F^{\gamma^*\gamma^*\pi^0}_\text{pert}
  \left(Q^2,-s-i\varepsilon\right)
  \right]
\end{eqnarray}
and
$s =\bar{x}Q^2/x$.
Then, one can express the TFF for one highly virtual and one almost
real photon as
\begin{eqnarray}
  Q^2 F^{\gamma^*\gamma\pi}\left(Q^2\right)
=
  \frac{\sqrt{2}}{3}f_\pi
  \left[
        \frac{Q^2}{m_{\rho}^2}
        \int_{x_{0}}^{1}
        \exp\left(
                  \frac{m_{\rho}^2-Q^2\bar{x}/x}{M^2}
            \right)
         ~\bar{\rho}(Q^2,x)
  \frac{dx}{x}
  + \! \int_{0}^{x_0} \bar{\rho}(Q^2,x)
        \frac{dx}{\bar{x}}
  \right] \, ,
\label{eq:LCSR-FQq}
\end{eqnarray}
where
$x_0 = Q^2/\left(Q^2+s_0\right)$
with $s_0\simeq 1.5$~GeV$^2$
and $M^2$ is the (auxiliary) Borel parameter.
The LCSR method has been applied to the pion-photon TFF in several
papers, see for instance,
\cite{Khodjamirian:1997tk, Schmedding:1999ap, Bakulev:2002uc, Bakulev:2003cs,
Bakulev:2005cp, Mikhailov:2009kf, Agaev:2010aq, Agaev:2012tm,
Bakulev:2011rp, Bakulev:2012nh, Stefanis:2012yw, Mikhailov:2016klg}.

On the experimental side, there are several sets of data from low
$Q^2$ values up to almost 40~GeV$^2$ taken at single-tag experiments.
The increase of the data above 10~GeV$^2$, observed by the
\textit{BABAR} Collaboration \cite{Aubert:2009mc}, has not been
confirmed by the more recent measurements of the Belle experiment
\cite{Uehara:2012ag}.
In the intermediate range of momenta $[5-9]$~GeV$^2$, both data sets
are compatible with each other and agree with the previous CLEO data
\cite{Gronberg:1997fj} but bear smaller statistical errors.
Finally, in the low-$Q^2$ regime below 5~GeV$^2$ down to 1~GeV$^2$
one can take recourse only to the CLEO data and the older CELLO data
\cite{Behrend:1990sr} which, however, have rather large uncertainties.
The BESIII Collaboration have announced high-precision data around
3~GeV$^2$ in the next future.\footnote{A. Denig, this conference.}
For the time being, they have only released simulated data in the
interval
$Q^2\in[0.5-3]$~GeV$^2$ \cite{Denig:2014mma}.
This ``data'' set merely serves to effect the expected small size of
the experimental errors in the event analysis.
This not withstanding, a higher accuracy and precision of experimental
data demands more reliable theoretical predictions and better control
of their uncertainties.
It is just this issue to which this study is devoted, while for the
full-fledged analysis we refer to \cite{Mikhailov:2016klg}.
In the next section we will expand the above exposition and consider
its main ingredients in more technical detail (Sec.\ \ref{sec:theory}).
Our main results and predictions will be given in Sec.\
\ref{sec:results} followed by our conclusions in Sec.\ \ref{sec:concl}.

\section{Theoretical framework}
\label{sec:theory}
\begin{figure}[h]                                                      
\centering\vspace*{5mm}
\sidecaption
\includegraphics[width=8cm,clip]{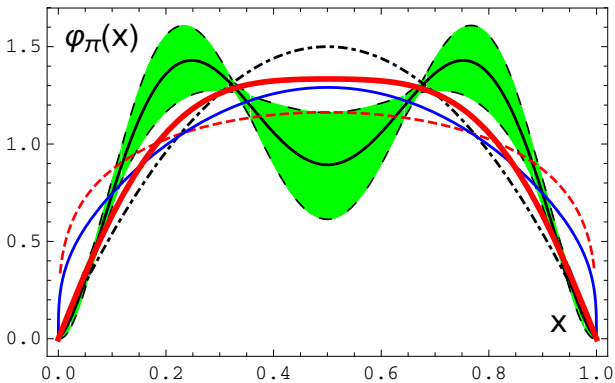} 
\caption{Various pion DAs at the momentum scale $\mu=2$~GeV.
The shaded band in green color displays the bimodal DAs derived in
\cite{Bakulev:2001pa} with QCD sum rules which employ nonlocal
condensates with a quark virtuality $\lambda_{q}^2 =0.4$~GeV$^2$.
The thick solid red line shows the platykurtic DA, obtained in
\cite{Stefanis:2014nla} using the same method but with the value
$\lambda_{q}^2 =0.45$~GeV$^2$.
The solid blue line just below it and the broad dashed curve in red
represent the results found within the DSE approach
\cite{Chang:2013pq} ---
termed DSE-DB and DSE-RL, respectively
(see text and Table \ref{tab:pion-DAs}).
The dashed-dotted black line is the asymptotic DA.}
\label{fig:pion-DA-1GeV}
\end{figure}

In continuation of the previous considerations, we now focus our
attention to the spectral density, starting with its twist
decomposition \cite{Khodjamirian:1997tk}:
\begin{eqnarray}
  \rho^{\text{pert}}(Q^{2},s)
& = &
   \rho_{\text{tw-2}}
  +\rho_{\text{tw-4}}
  +\rho_{\text{tw-6}}
  +\ldots\, .
\label{eq:rho-twists}
\end{eqnarray}
The next step is to expand the twist-two part of $\bar{\rho}(Q^{2},x)$
in (\ref{eq:LCSR-FQq}) in terms of the partial spectral densities
$\bar{\rho}_n$ which are associated with the eigenfunctions $\psi_n$
of the ERBL evolution equation.
The set $\{\psi_{n}\}$ represents a conformal basis for the expansion
of the pion DA which reads
\begin{equation}
  \varphi_{\pi}^{(2)}(x,\mu^{2})
=\psi_{0}(x)
  + \sum_{n=2,4, \ldots}^{\infty} a_{n}(\mu^{2}) \psi_{n}(x) \, ,
\label{eq:gegen-exp}
\end{equation}
where the conformal coefficients $a_{n}$ encode the
nonperturbative properties of the DA, while
the lowest-order term of this expansion,
$\psi_{0}(x)=\varphi_{\pi}^{\rm asy}=6x\bar{x}$
represents the asymptotic (asy) DA (dashed-dotted line in
Fig.\ \ref{fig:pion-DA-1GeV}).

The sum over partial densities then becomes
\begin{eqnarray}
  \bar{\rho}\left(Q^2,x\right)
=
  \sum_{n=0,2,4,\ldots}a_{n}\left(Q^2\right)
    \bar{\rho}_{n}\left(Q^2,x\right)
  + \bar{\rho}_\text{tw-4}\left(Q^2,x\right)
  + \bar{\rho}_\text{tw-6}\left(Q^2,x\right) \, ,
\label{eq:rho-bar-15}
\end{eqnarray}
where
\begin{eqnarray}
  \bar{\rho}_{n}\left(Q^2,x\right)
\!\!\!&\!\!=\!\! &\!\!\!
    \bar{\rho}_{n}^{(0)}(x)
    + a_{s}\bar{\rho}_{n}^{(1)}(Q^2,x)
    + a_{s}^{2}\bar{\rho}_{n}^{(2)}(Q^2,x)
    + \ldots, \nonumber \\
  \bar{\rho}_{n}^{(0)}(x)
&=&
  \psi_n(x);~\,
  a_s
=
  a_{s}(Q^2) \, ,
\label{eq:rho-n}
\end{eqnarray}
with the explicit expressions for the various terms being given in
Appendix B of Ref.\ \cite{Mikhailov:2016klg}.

The other two contributions to the spectral density of higher twist
comprise the twist-four and the twist-six terms.
The twist-four term is given by \cite{Khodjamirian:1997tk}
\begin{equation}
  \bar{\rho}_{\text{tw-4}}(Q^2,x)
=
  \frac{\delta^2_\text{tw-4}(Q^2)}{Q^2}
  x\frac{d}{dx}\varphi^{(4)}_\pi(x)\, ,
\label{eq:rho-tw-4}
\end{equation}
where the twist-four coupling parameter is defined by
$
 \delta^2_\text{tw-4}
\approx
 (1/2)\lambda^{2}_{q}
=
 (1/2)\left(0.4 \pm 0.05 \right)$~GeV$^2$
at $Q^2 \approx 1$ GeV$^2$ \cite{Bakulev:2002uc}
and $\lambda^{2}_{q}$ is the average virtuality
of vacuum quarks \cite{Mikhailov:1986be}.
In our actual calculations the twist-four pion DA is approximated
by its asymptotic form \cite{Khodjamirian:1997tk, Agaev:2010aq}
\begin{equation}
\varphi_{\pi}^{(4)}(x)= \frac{80}{3}x^2(1-x)^2 \, .
\label{eq:tw-4-DA}
\end{equation}
The twist-six term in Eq.\ (\ref{eq:rho-bar-15}),
$\bar{\rho}_{\text{tw-6}}(Q^{2},x)
=
 (Q^2+s)\rho_{\text{tw-6}}(Q^2,s)
$,
was first computed in \cite{Agaev:2010aq} and was recently confirmed by
an independent calculation in \cite{Mikhailov:2016klg}.
It can be expressed as follows
\begin{eqnarray}
 \label{eq:tw-6}
    \bar{\rho}_\text{tw-6}(Q^{2}\!,x)
=
    8\pi \frac{C_\text{F}}{N_c}
    \frac{ \alpha_s\langle\bar{q} q\rangle^2}{f_\pi^2}\frac{x}{Q^4}
    \left[
        \!-\!
        \left[\frac{1}{1-x}\right]_+
        \!+\!\left(2\delta(\bar{x})-4 x\right)\!+\!
        \left(
         3x+2x\log{x}
        \!+\!
        2x\log{}\bar{x}
        \right)
    \right]
\end{eqnarray}
with
$\alpha_s=0.5$, $C_\text{F}=4/3$, $N_c=3$, and $
 \langle \bar{q} q\rangle^2
=
 \left(0.242 \pm 0.01 \right)^6$~GeV$^6$ \cite{Gelhausen:2013wia}.

The key virtue of the LCSR method relative to the factorization
approach of perturbative QCD is that it offers the possibility to
include into the TFF the higher eigenfunctions $\psi_{n>0}(x)$ in
a successive way in accordance with the increase of the
momentum $Q^2$.
Thus, one can evaluate the TFF
\begin{eqnarray}
  Q^2F^{\gamma^*\gamma\pi^0}(Q^2)
=
    \underbrace{F_0(Q^2)
  + \sum_{n} a_n(Q^2)F_n(Q^2)}_{\mbox{\scriptsize Tw-2}}
  + F^{(4)}(Q^2)
  + F^{(6)}(Q^2) \, ,
\label{eq:TFF-full}
\end{eqnarray}
where the underbraced terms constitute the leading twist-two
contribution, sequentially, i.e., by including with increasing
$Q^2$ more and more terms of the conformal expansion over
$\psi_n(x)$.
In contrast, the lowest-order leading-twist contribution obtained
in perturbative QCD is
\begin{equation}
  Q^2F_{n}^{\rm pQCD}(Q^2)
=
  \int_{0}^{1}\psi_n(x) \frac{dx}{\bar{x}}
=
  3~~~~~\text{for}~\forall ~n
\label{eq:TFF-pQCD}
\end{equation}
and therefore all terms of the conformal
expansion contribute at once with the effect that $F_n(Q^2)$ does
not vary with $Q^2$
(see \cite{Mikhailov:2016klg} for an illustration of this procedure).
An important observation is that at low momenta
$Q^2\simeq 1$~GeV$^2$
mainly the zeroth-order contribution $F_0$ survives, while the
spectral decomposition of the pion DA is less important.

\begin{figure}[h]                                                      
\centering
\includegraphics[width=9cm,clip]{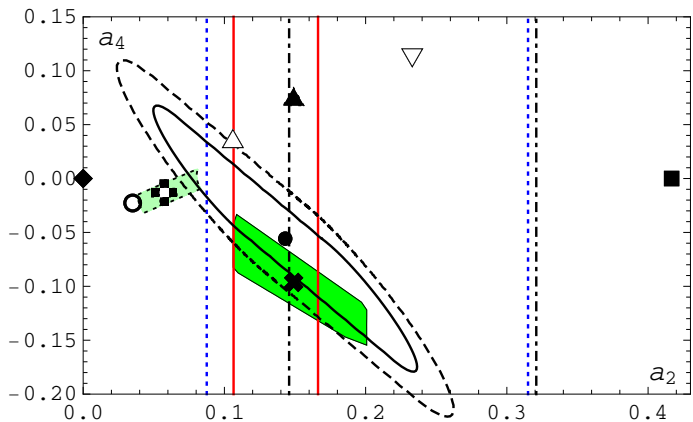}
\caption{Locations of various pion DAs or of its
2D projections in the space of the conformal coefficients
$a_2$ and $a_4$ at the scale $\mu=2$~GeV.
The designations are given in Table \ref{tab:pion-DAs}, while
theoretical explanations are provided in the text.}
\label{fig:a_2-a_4-space}
\end{figure}

Let us now turn our attention to the major nonperturbative input of
the TFF, notably, the pion DA.
This important quantity is not measurable by itself.
Therefore, one has to rely upon its moments
\begin{equation}
  \langle \xi^{N} \rangle_{\pi}
\equiv
  \int_{0}^{1} \varphi_{\pi}^{(2)}(x,\mu^2) (x-\bar{x})^{N}\, dx ,
\label{eq:moments}
\end{equation}
where $\xi = x - \bar{x}$,
from which the conformal coefficients $a_n$ in
Eq.\ (\ref{eq:gegen-exp}) can be determined
(see, for instance, \cite{Chernyak:1983ej,Stefanis:1999wy}
for reviews).
For the sake of definiteness, the discussion of the theoretical
uncertainties affecting the calculation of the pion-photon TFF is based
in our analysis on the pion DAs obtained with the help of QCD sum rules
and nonlocal condensates, starting with the paper by the late Bakulev
and two of us (Mikhailov and Stefanis) in \cite{Bakulev:2001pa}.
This set of DAs, termed BMS,
(green band in Fig.\ \ref{fig:pion-DA-1GeV})
was determined by using the mean value of the vacuum quark virtuality
$\lambda_{q}^{2}=0.4$~GeV$^2$.
All DAs belonging to this set are bimodal distributions with a more or
less pronounced dip at $x=1/2$ and strongly suppressed tails at
$x=0,1$.
Their domain in the parameter space spanned by the Gegenbauer
coefficients $a_2$ and $a_4$ is shown in the form of a slanted
green rectangle in Fig.\ \ref{fig:a_2-a_4-space},
with the symbol \ding{54} denoting the BMS model DA.
More recently, another DA was derived within the same approach
\cite{Stefanis:2014nla, Stefanis:2014yha, Stefanis:2015qha},
but using the slightly, but still admissible, vacuum quark virtuality
$\lambda_{q}^{2}=0.45$~GeV$^2$.
This DA is unimodal with a broad peak at $x=1/2$ but still exhibiting
strong suppression of the endpoint regions at $x=0,1$, thus giving rise
to a platykurtic profile
(solid red line in Fig.\ \ref{fig:pion-DA-1GeV}).
A dedicated investigation of the endpoint regions of the
twist-two pion DA was performed in \cite{Mikhailov:2010ud} to which we
refer for further reading.
Note that arguments for the endpoint suppression of the pion DA were
given even earlier \cite{Stefanis:1998dg,Stefanis:2000vd,Choi:2007yu}
in the context of quantum fluctuations of the QCD vacuum and
the appearance of fermionic zero modes.
The graphical representation of the domain of the platykurtic DAs is
given by the small slanted rectangle in light-green color in
Fig.\ \ref{fig:a_2-a_4-space}, with the platykurtic DA
\cite{Stefanis:2014nla} being denoted by the symbol \ding{60}.
Both types of DAs employ only the first nontrivial coefficients $a_2$
and $a_4$, with the higher ones up to order 10 being determined and
found to be close to zero but bearing large uncertainties.
Their values were selected to fit best all moments
$\langle \xi^N\rangle$ with $N=2,4,\ldots,10$ with the mentioned
$\lambda_{q}^{2}$ values \cite{Bakulev:2001pa,Stefanis:2015qha}.
Note that the predictions for the pion-photon TFF computed with the
platykurtic DA are very close to the BMS ones
\cite{Stefanis:2014nla, Mikhailov:2016klg}.
The locations of several other DA models or their 2D projections have
been included in Fig.\ \ref{fig:a_2-a_4-space} as well
(designations in Table \ref{tab:pion-DAs}).
These models are compared with the $1\sigma$ and $2\sigma$ error
ellipses, centered around {\footnotesize\ding{108}}, which were
obtained by means of LCSRs from the combined analysis of the
CELLO \cite{Behrend:1990sr}, CLEO \cite{Gronberg:1997fj},
\textit{BABAR}($\leqslant 9$~GeV$^2$) \cite{Aubert:2009mc}, and Belle
\cite{Uehara:2012ag} data.
The displayed models refer to the scale
$Q^2=4$~GeV$^2$ after NLO ERBL evolution to that scale, provided the
original normalization scale was lower.
To complete the picture, also the constraints on $a_2$ obtained by
three different lattice collaborations are also shown in this
figure in terms of vertical lines (adding the various errors
linearly):
solid red lines ---
$a_2=0.1364\pm 0.0299$
\cite{Braun:2015axa}\footnote{The subtleties related
to the continuum limit in extracting $a_2$ from
$\langle \xi^2 \rangle$ have been discussed in
\cite{Stefanis:2015qha}.},
dashed blue lines --- $0.201(114)$ \cite{Braun:2006dg},
dashed-dotted lines --- $0.233(87)$ (result for the
larger lattice)
\cite{Arthur:2010xf}.
The values of the Gegenbauer coefficients of the pion DAs shown in
Fig.\ \ref{fig:a_2-a_4-space} are collected in
Table \ref{tab:pion-DAs}.
We also include the value of the inverse moment
$
 \langle 1/x \rangle_\pi = \int_{0}^{1} dx x^{-1}
 \varphi_{\pi}^{(2)}(x)
=
 3(1+a_2 + a_4 + \ldots )
=
 3/(\sqrt{2}f_{\pi})Q^2 F_{\gamma\pi}^{(2)}(Q^2)
$
which gives a crude estimate of the magnitude of the TFF at the
Twist-2 level of accuracy.
For endpoint-enhanced DAs, this quantity receives additional
contributions from the omitted (positive) higher-order coefficients as
one sees from the numbers in parentheses which take into account all
terms up to $a_{12}$ \cite{Raya:2015gva}\footnote{All coefficients
up to order 40 are positive numbers --- P. Tandy, this conference.}.
\begin{table}[h]                                                       
\centering
\caption{Conformal coefficients $a_2$ and $a_4$, and inverse moment
$\langle 1/x\rangle$
at the scale $\mu^2=4$~GeV$^2$ related to the pion DAs shown in
Fig.\ \ref{fig:a_2-a_4-space}.
NLO evolution is applied if the normalization scale was lower.}
\label{tab:pion-DAs}
\begin{tabular}{llll}
\hline
Pion DA                                                               & $a_2$                      & $a_4$                       & $\langle 1/x \rangle_\pi$ \\\hline
BMS rectangle \cite{Bakulev:2001pa, Bakulev:2011rp}                   & [0.11, 0.20]               & $[-0.15, -0.03]$            & $[2.88, 3.51]$            \\
small slanted green rectangle \cite{Stefanis:2015qha}                 & [0.04, 0.08]               & $[-0.03, 0.01]$             & $[3.03, 3.27]$            \\
BMS model \cite{Bakulev:2001pa, Mikhailov:2016klg} \ding{54}          & $0.149_{-0.043}^{+0.052}$  & $-0.096_{-0.058}^{+0.063}$  & $3.159_{-0.09}^{+0.09}$   \\
platykurtic model \cite{Stefanis:2014nla, Stefanis:2015qha} \ding{60} & $0.057^{+0.024}_{-0.019}$  & $-0.013^{+0.022}_{-0.019}$  & $3.132_{-0.10}^{+0.14}$   \\
center of $\sigma$ ellipses \cite{Bakulev:2011rp} {\footnotesize\ding{108}}& 0.143                 & $-0.056$                    & 3.261                     \\
DSE-DB \cite{Chang:2013pq} \ding{115}                                 & 0.149                      & 0.076                       & 3.675 (3.835)             \\
DSE-RL \cite{Chang:2013pq} $\bigtriangledown$                         & 0.233                      & 0.112                       & 4.035 (4.527)             \\
AdS/QCD \cite{Brodsky:2011yv} $\bigtriangleup$                        & 0.107                      & 0.038                       & 3.435                     \\
Light-Front model \cite{Choi:2014ifm} $\bigcirc$                      & 0.035                      & $-0.023$                    & 3.036                     \\
CZ model \cite{Chernyak:1983ej} \ding{110}                            & 0.412                      & 0                           & 4.236                     \\
asymptotic \ding{117}                                                 & 0                          & 0                           & 3                         \\\hline
\end{tabular}
\end{table}

\section{Results and predictions}
\label{sec:results}
The theoretical scheme described in the previous section can now be
used to derive predictions for the $F^{\gamma*\gamma\pi^0}(Q^2)$
TFF from small to large values of the momentum transfer $Q^2$.
We have presented such predictions before focusing attention to the
large-$Q^2$ regime
\cite{Bakulev:2011rp, Stefanis:2012yw, Bakulev:2012nh}.
More recently, we addressed the low-to-middle range
$Q^2\leq 5$~GeV$^2$ motivated by the simulated BESIII-data in this
domain \cite{Mikhailov:2016klg}.
In order to match the promised high statistical accuracy of these
data, we attempted i) to improve the knowledge of the theoretical
ingredients of the analysis and ii) to estimate more reliably the
associated theoretical uncertainties.
With regard to the first issue, let us display the various terms
entering Eq.\ (\ref{eq:hard-scat-ampl}) for the hard-scattering
amplitude $T$ up to the contribution proportional to $a_s^2$:
\begin{subequations}
\label{eq:T}
\begin{eqnarray}
  T_{\rm LO}
& = &
  T_0,
\\
  T_{\rm NLO}
& = & C_{\rm F}~
  T_0 \otimes \left[\mathcal{T}^{(1)}+
                      L~  V_{+}^{(0)}
              \right],
\label{eq:NLO}
\\
  T_{\rm NNLO}
& = &C_{\rm F}~
  T_0 \otimes \left[
                     {\cal T}^{(2)} + L~ V_{+}^{(1)}/C_{\rm F}
                     - L~ \beta_{0}  {\cal T}^{(1)} \right.  \nonumber \\
                     && - \left.
                     \frac{L^2}{2}~ \beta_{0} V_{+}^{(0)}
                     + \frac{L^2}{2}~ C_{\rm F} V_{+}^{(0)}\otimes V_{+}^{(0)}\right.
\nonumber \\
                     && + \left.
                     L~ C_{\rm F} {\cal T}^{(1)}\otimes V_{+}^{(0)}
              \right] \, .
\label{eq:hard-scat-series}
\end{eqnarray}
\end{subequations}
Note that here and below the convenient abbreviation \cite{Melic:2002ij}
$
 L
\equiv
 \ln\left[\left(Q^2y+q^2\bar{y}\right)/\mu^2_\text{F}\right]
$
is used.

The NLO term $T_\text{NLO}$ is completely known
\cite{Bakulev:2002uc, Melic:2002ij}.
Consider now the NNLO term $T_\text{NNLO}$ which can be cast in the
form (see \cite{Mikhailov:2016klg})
\begin{eqnarray}
  T_{\rm NNLO}
= C_{\rm F} T_0 \otimes \!
    \left[\beta_0 T_\beta
  + T_{\Delta V}
  + T_L+  \mathcal{T}^{(2)}_c \right]\, ,
\label{eq:NNLO-beta}
\end{eqnarray}
where
\begin{subequations}
 \label{eq:T-elements}
\begin{eqnarray}
   T_{\Delta V}
& = &
 L \Delta V^{(1)}_+,
\label{eq:hard-scat.nnlo-dv} \\
  T_L
& = &
 C_{\rm F} L \left(
           \frac{L}{2} V_{+}^{(0)}\otimes V_{+}^{(0)}
+ \mathcal{T}^{(1)}\otimes V_{+}^{(0)}
             \right)
                     \, .
\label{eq:hard-scat-nnlo}
\end{eqnarray}
\end{subequations}
The explicit expressions for the above terms can be found in
Appendix A of Ref.\ \cite{Mikhailov:2016klg}.
What is worth emphasizing here is that the terms $T_{\Delta V}$ and
$T_L$
have been calculated for the first time in \cite{Mikhailov:2016klg} and
are given there in Appendix B.
These two additional terms give a small contribution to the TFF as we
will discuss later.
Finally, the term $T_c^{(2)}$ is the remaining term at NNLO which
is still unknown.
On the nonperturbative side, we recalculated term-by-term in
\cite{Mikhailov:2016klg} the twist-six correction to the TFF and
confirmed the original computation in \cite{Agaev:2010aq}.
These improvements increase the reliability of our theoretical
predictions and gives us better control over the uncertainties
associated with them.

The main theoretical uncertainties originate from three different
sources, which can be ordered according to their twist:
\begin{itemize}
\item Tw-2: Variation of the pion DA in terms of the conformal
coefficients $a_n(Q^2)$ --- custom to the method used.
\item Tw-2: Uncalculated NNLO term $T_c$ in the TFF.
\item Tw-4: Value of $\lambda_{q}^{2}$
(related to $\delta^2_\text{tw-4}$)
entering the calculation of the DAs in the approach with nonlocal
condensates and independently also the computation of the TFF
via the spectral density $\bar{\rho}_{\rm tw-4}(Q^2,x)$ in LCSRs.
\item Tw-6: Value of $\alpha_s\langle \bar{q}q\rangle^2$ parameterizing
the spectral density $\bar{\rho}_{\rm tw-6}(Q^2,x)$.
\end{itemize}
In addition, there are theoretical uncertainties related to auxiliary
parameters employed in the LCSRs.
These are the Borel parameter $M^2$ and the parametrization procedure
of the $(\omega)\rho$-meson resonance on the right-hand side of
Eq.\ (\ref{eq:LCSR-FQq}).
Finally, the calculated TFF depends on the small virtuality of the
quasi-real photon whose value is fixed by each particular experiment.
In our work we employ for the estimation of this effect the value
$q^2=0.04$~GeV$^2$ which is related to the Belle
experiment.\footnote{S.\ Uehara, private communication.}
The detailed estimates in percentage of all these uncertainties
at $Q^2=3$~GeV$^2$ were tabulated in \cite{Mikhailov:2016klg} using
$M^2\in[0.7-1.0]$~GeV$^2$ as in \cite{Bakulev:2011rp, Bakulev:2012nh},
but also the value $M^2=1.5$~GeV$^2$ employed in
\cite{Agaev:2010aq, Agaev:2012tm}.

\begin{figure}[h]                                                      
\centering
\includegraphics[width=8cm,clip]{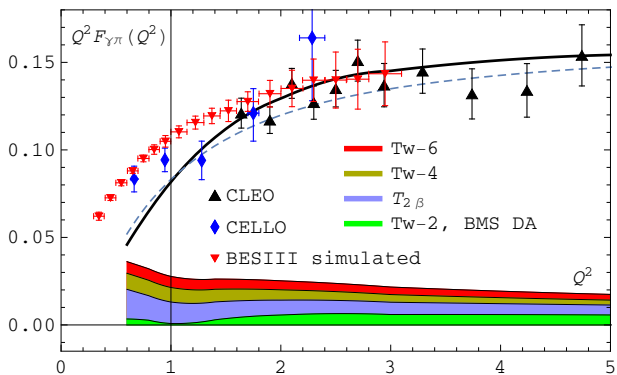}
\caption{Predictions for the pion-photon TFF within our LCSR method
at low-$Q^2$ in comparison with the existing data of
CELLO \cite{Behrend:1990sr} and CLEO \cite{Gronberg:1997fj},
and the simulated data of BESIII \cite{Denig:2014mma}.
The solid black line shows the result for the BMS DA and the
dashed blue line below it the analogous result obtained with the
platykurtic DA.
Only the key theoretical errors are displayed in the form of colored
strips as indicated (see text).}
\label{fig:TFF-pred}
\end{figure}

Here, we only illustrate in Fig.\ \ref{fig:TFF-pred} the select main
theoretical uncertainties by displaying $Q^2F_{\gamma\pi}(Q^2)$ in the
momentum range $\leq 5$~GeV$^2$ for the BMS DA (solid black line) and
for the platykurtic one (dashed blue line).
The amount of these uncertainties is effected in terms
of colored strips, as indicated in the figure.
The influence of the variation of the auxiliary quantities, mentioned
above, and the impact of a finite virtuality of the quasi-real photon
can be found in our encompassing analysis in \cite{Mikhailov:2016klg}.
Note that the reliability of our theoretical predictions below
$\sim 1$~GeV$^2$ becomes insufficient.

Some more comments are here in order.
Our comparison in Fig.\ \ref{fig:a_2-a_4-space} also includes
the $(a_2,a_4)$ projections of the two broad unimodal pion DAs obtained
from Dyson-Schwinger equations (DSE) \cite{Chang:2013pq}.
The DA termed DSE-RL was derived by using the rainbow-ladder truncation
(symbol $\bigtriangledown$), while its more advanced version --- coined
DSE-DB --- employs the most improved kernel to express Dynamical Chiral
Symmetry Breaking (DCSB) (symbol \ding{115}) --- called
``DB truncation''.
As one sees from Table \ref{tab:pion-DAs}, the DSE-DB $\pi$ DA
leads at $\mu^2=4$~GeV$^2$ to the same $a_2$ value 0.149 as the BMS DA.
This value is supported by the latest lattice simulation in
\cite{Braun:2015axa} but is, in contrast to the BMS DA, outside the
$1\sigma$ (solid line) and $2\sigma$ (dashed line) error ellipses
obtained with a LCSR-fit to the CELLO, CLEO,
\textit{BABAR}($\leqslant 9$~GeV$^2$), and Belle data
(see Fig.\ \ref{fig:a_2-a_4-space})
because $a_4$ is sizeable and positive.
Remarkably, the authors of \cite{Raya:2015gva} obtain with the DSE-DB
pion DA a TFF prediction within their framework which agrees well
with all these data and thus belongs to the green band of
predictions described in \cite{Bakulev:2012nh}.
Last but not least: the similarly broad DA $(8/\pi)\sqrt{x\bar{x}}$,
based on the AdS/QCD and light-front holography \cite{Brodsky:2011yv}
(its 2D projection is denoted by the symbol $\bigtriangleup$ in
Fig.\ \ref{fig:a_2-a_4-space}) turns out to be just inside the
$2\sigma$ error ellipse of the above sets of experimental data yielding
a TFF prediction \cite{Brodsky:2011yv,Brodsky:2011xx} which belongs to
the green band in \cite{Bakulev:2012nh} as well.

\section{Conclusions}
\label{sec:concl}
In this paper we have described a theoretical approach based on LCSRs
to handle the calculation of the pion-photon TFF
$F^{\gamma^*\gamma\pi^0}(Q^2)$
including its main theoretical uncertainties which originate
from different ingredients of the approach.
The key observations pertaining to the reliability of our predictions
and the role of the involved uncertainties can be summarized as
follows.
\begin{itemize}
\item
The unknown NNLO term $T_c$ of the radiative corrections generates
the largest uncertainty.
\item
Most of the theoretical uncertainties are correlated at different
values of $Q^2$.
\item
At low momenta, the TFF is more sensitive to the variations of the
higher-twist contributions Tw-4 and Tw-6 and less sensitive to the
values of the conformal parameters $a_n$ describing the pion DA.
Indeed, in the vicinity of $Q^2=1$~GeV$^2$ the sensitivity
to $a_n$ disappears completely as one observes from the green strip
at the bottom of Fig.\ \ref{fig:TFF-pred}.

\item
The platykurtic pion DA yields a TFF prediction close to the
result obtained with the original bimodal BMS DA.
The common feature of these DAs is the suppression of the endpoints.
Both types of DAs yield predictions which agree rather well with the
available experimental data in the range $1\leq Q^2\leq 5$~GeV$^2$
and also all higher ones compatible with QCD scaling
\cite{Bakulev:2012nh, Stefanis:2014nla}.
\end{itemize}

It is interesting to extend the above discussion a bit further and
mention that also the DA determined in Ref.\ \cite{Choi:2014ifm} within
a light-front quark model has a platykurtic shape
(symbol $\bigcirc$ in Fig.\ \ref{fig:a_2-a_4-space}) and reproduces
all data on $F^{\gamma^*\gamma\pi^0}(Q^2)$, i.e., belongs to the green
band of predictions in the classification scheme of
\cite{Bakulev:2012nh}.
In the same context we note that a recent simultaneous analysis of the
CLEO \cite{Gronberg:1997fj} and Belle \cite{Uehara:2012ag} data in
Ref.\ \cite{Zhong:2015nxa} favors a platykurtic DA profile as well ---
see \cite{Stefanis:2015qha} for a more detailed comparison.
Moreover, a brand-new analysis of the TFF finds best agreement with all
sets of existing data except those of \textit{BABAR} above 10~GeV$^2$
using a spin-improved holographic pion twist-two DA with platykurtic
profile \cite{Ahmady:2016ufq}.

We look ahead for the high-precision data by the BESIII Collaboration
in the near future and the high-$Q^2$ data on two-photon physics to be
measured by the BelleII experiment at the end of this decade.

\begin{acknowledgement}
This work was partially supported by the Heisenberg--Landau Program
(Grant 2016), and the Russian Foundation for Basic Research
under Grants No.\ 14-01-00647, No.\ 15-52-04023.
A.P. has been supported by the ``Chinese Academy of Sciences
President's International Fellowship Initiative'' under
Grant No.\ 2016PM053, and NSFC Grants No.\ 11650110431 and
No.\ 11575254.
\end{acknowledgement}

\bibliographystyle{woc}

\begin{thebibliography}{46}

\bibitem{Efremov:1979qk}
A.V. Efremov, A.V. Radyushkin, Phys. Lett. \textbf{B94}, 245 (1980)

\bibitem{Efremov:1978rn}
A.V. Efremov, A.V. Radyushkin, Theor. Math. Phys. \textbf{42}, 97 (1980)

\bibitem{Lepage:1979zb}
G.P. Lepage, S.J. Brodsky, Phys. Lett. \textbf{B87}, 359 (1979)

\bibitem{Lepage:1980fj}
G.P. Lepage, S.J. Brodsky, Phys. Rev. \textbf{D22}, 2157 (1980)

\bibitem{Brodsky:1981rp}
S.J. Brodsky, G.P. Lepage, Phys. Rev. \textbf{D24}, 1808 (1981)

\bibitem{Melic:2002ij}
B.~Meli\'c, D.~M{\"u}ller, K.~Passek-Kumeri\v{c}ki, Phys. Rev. \textbf{D68},
  014013 (2003), \texttt{hep-ph/0212346}

\bibitem{Bakulev:2002uc}
A.P. Bakulev, S.V. Mikhailov, N.G. Stefanis, Phys. Rev. \textbf{D67}, 074012
  (2003), \texttt{hep-ph/0212250}

\bibitem{Mikhailov:2016klg}
S.V. Mikhailov, A.V. Pimikov, N.G. Stefanis, Phys. Rev. \textbf{D93}, 114018
  (2016), \texttt{1604.06391}

\bibitem{Balitsky:1989ry}
I.I. Balitsky, V.M. Braun, A.V. Kolesnichenko, Nucl. Phys. \textbf{B312}, 509
  (1989)

\bibitem{Khodjamirian:1997tk}
A.~Khodjamirian, Eur. Phys. J. \textbf{C6}, 477 (1999), \texttt{hep-ph/9712451}

\bibitem{Schmedding:1999ap}
A.~Schmedding, O.I. Yakovlev, Phys. Rev. \textbf{D62}, 116002 (2000),
  \texttt{hep-ph/9905392}

\bibitem{Bakulev:2003cs}
A.P. Bakulev, S.V. Mikhailov, N.G. Stefanis, Phys. Lett. \textbf{B578}, 91
  (2004), \texttt{hep-ph/0303039}

\bibitem{Bakulev:2005cp}
A.P. Bakulev, S.V. Mikhailov, N.G. Stefanis, Phys. Rev. \textbf{D73}, 056002
  (2006), \texttt{hep-ph/0512119}

\bibitem{Mikhailov:2009kf}
S.V. Mikhailov, N.G. Stefanis, Nucl. Phys. \textbf{B821}, 291 (2009),
  \texttt{0905.4004}

\bibitem{Agaev:2010aq}
S.S. Agaev, V.M. Braun, N.~Offen, F.A. Porkert, Phys. Rev. \textbf{D83}, 054020
  (2011), \texttt{1012.4671}

\bibitem{Agaev:2012tm}
S.S. Agaev, V.M. Braun, N.~Offen, F.A. Porkert, Phys. Rev. \textbf{D86}, 077504
  (2012), \texttt{1206.3968}

\bibitem{Bakulev:2011rp}
A.P. Bakulev, S.V. Mikhailov, A.V. Pimikov, N.G. Stefanis, Phys. Rev.
  \textbf{D84}, 034014 (2011), \texttt{1105.2753}

\bibitem{Bakulev:2012nh}
A.P. Bakulev, S.V. Mikhailov, A.V. Pimikov, N.G. Stefanis, Phys. Rev.
  \textbf{D86}, 031501 (2012), \texttt{1205.3770}

\bibitem{Stefanis:2012yw}
N.G. Stefanis, A.P. Bakulev, S.V. Mikhailov, A.V. Pimikov, Phys. Rev.
  \textbf{D87}, 094025 (2013), \texttt{1202.1781}

\bibitem{Aubert:2009mc}
B.~Aubert et~al. (BaBar), Phys. Rev. \textbf{D80}, 052002 (2009),
  \texttt{0905.4778}

\bibitem{Uehara:2012ag}
S.~Uehara et~al. (Belle), Phys. Rev. \textbf{D86}, 092007 (2012),
  \texttt{1205.3249}

\bibitem{Gronberg:1997fj}
J.~Gronberg et~al. (CLEO), Phys. Rev. \textbf{D57}, 33 (1998),
  \texttt{hep-ex/9707031}

\bibitem{Behrend:1990sr}
H.J. Behrend et~al. (CELLO), Z. Phys. \textbf{C49}, 401 (1991)

\bibitem{Denig:2014mma}
A.~Denig (BESIII), Nucl. Part. Phys. Proc. \textbf{260}, 79 (2015),
  \texttt{1412.2951}

\bibitem{Bakulev:2001pa}
A.P. Bakulev, S.V. Mikhailov, N.G. Stefanis, Phys. Lett. \textbf{B508}, 279
  (2001), [Erratum: Phys. Lett. B590, 309 (2004)], \texttt{hep-ph/0103119}

\bibitem{Stefanis:2014nla}
N.G. Stefanis, Phys. Lett. \textbf{B738}, 483 (2014), \texttt{1405.0959}

\bibitem{Chang:2013pq}
L.~Chang, I.C. Cloet, J.J. Cobos-Martinez, C.D. Roberts, S.M. Schmidt, P.C.
  Tandy, Phys. Rev. Lett. \textbf{110}, 132001 (2013), \texttt{1301.0324}

\bibitem{Mikhailov:1986be}
S.V. Mikhailov, A.V. Radyushkin, JETP Lett. \textbf{43}, 712 (1986), [Pisma Zh.
  Eksp. Teor. Fiz.43,551(1986)]

\bibitem{Gelhausen:2013wia}
P.~Gelhausen, A.~Khodjamirian, A.A. Pivovarov, D.~Rosenthal, Phys. Rev.
  \textbf{D88}, 014015 (2013), [Erratum: Phys. Rev. D91, 099901 (2015)],
  \texttt{1305.5432}

\bibitem{Chernyak:1983ej}
V.L. Chernyak, A.R. Zhitnitsky, Phys. Rept. \textbf{112}, 173 (1984)

\bibitem{Stefanis:1999wy}
N.G. Stefanis, Eur. Phys. J. direct \textbf{C7}, 1 (1999),
  \texttt{hep-ph/9911375}

\bibitem{Stefanis:2014yha}
N.G. Stefanis, S.V. Mikhailov, A.V. Pimikov, Few Body Syst. \textbf{56}, 295
  (2015), \texttt{1411.0528}

\bibitem{Stefanis:2015qha}
N.G. Stefanis, A.V. Pimikov, Nucl. Phys. \textbf{A945}, 248 (2016),
  \texttt{1506.01302}

\bibitem{Mikhailov:2010ud}
S.V. Mikhailov, A.V. Pimikov, N.G. Stefanis, Phys. Rev. \textbf{D82}, 054020
  (2010), \texttt{1006.2936}

\bibitem{Stefanis:1998dg}
N.G. Stefanis, W.~Schroers, H.C. Kim, Phys. Lett. \textbf{B449}, 299 (1999),
  \texttt{hep-ph/9807298}

\bibitem{Stefanis:2000vd}
N.G. Stefanis, W.~Schroers, H.C. Kim, Eur. Phys. J. \textbf{C18}, 137 (2000),
  \texttt{hep-ph/0005218}

\bibitem{Choi:2007yu}
H.M. Choi, C.R. Ji, Phys. Rev. \textbf{D75}, 034019 (2007),
  \texttt{hep-ph/0701177}

\bibitem{Braun:2015axa}
V.M. Braun, S.~Collins, M.~G{\"o}ckeler, P.~P\'erez-Rubio, A.~Sch{\"a}fer, R.W.
  Schiel, A.~Sternbeck, Phys. Rev. \textbf{D92}, 014504 (2015),
  \texttt{1503.03656}

\bibitem{Braun:2006dg}
V.M. Braun et~al., Phys. Rev. \textbf{D74}, 074501 (2006),
  \texttt{hep-lat/0606012}

\bibitem{Arthur:2010xf}
R.~Arthur, P.A. Boyle, D.~Br{\"o}mmel, M.A. Donnellan, J.M. Flynn,
  A.~J{\"u}ttner, T.D. Rae, C.T.C. Sachrajda, Phys. Rev. \textbf{D83}, 074505
  (2011), \texttt{1011.5906}

\bibitem{Raya:2015gva}
K.~Raya, L.~Chang, A.~Bashir, J.J. Cobos-Martinez, L.X. Guti\'errez-Guerrero,
  C.D. Roberts, P.C. Tandy, Phys. Rev. \textbf{D93}, 074017 (2016),
  \texttt{1510.02799}

\bibitem{Brodsky:2011yv}
S.J. Brodsky, F.G. Cao, G.F. de~T\'eramond, Phys. Rev. \textbf{D84}, 033001
  (2011), \texttt{1104.3364}

\bibitem{Choi:2014ifm}
H.M. Choi, C.R. Ji, Phys. Rev. \textbf{D91}, 014018 (2015), \texttt{1412.2507}

\bibitem{Brodsky:2011xx}
S.J. Brodsky, F.G. Cao, G.F. de~T\'eramond, Phys. Rev. \textbf{D84}, 075012
  (2011), \texttt{1105.3999}

\bibitem{Zhong:2015nxa}
T.~Zhong, X.G. Wu, T.~Huang, Eur. Phys. J. \textbf{C76}, 390 (2016),
  \texttt{1510.06924}

\bibitem{Ahmady:2016ufq}
M.~Ahmady, F.~Chishtie, R.~Sandapen (2016), \texttt{1609.07024}

\end{thebibliography}

\end{document}